# Origin of Enhanced Thermal Resistance Near Nanoscale Hotspots: Insights from Full-Dispersion-Resolved Phonon Transport in Silicon

Jae Sik Jin [a]

*Department of Mechanical Design, Chosun College of Science & Technology, Gwangju 61453, Republic of Korea*

**Abstract** Phonon transport near nanoscale hotspots (NHs) critically determines heat dissipation in advanced electronic devices. The prevailing understanding is that the enhanced thermal resistance (TR) observed in NHs originates from long mean free path (MFP) phonons, whose MFPs are much larger than the hotspot size, thereby limiting their ability to recognize hotspots and transport heat effectively. In this study, we revisit this problem by employing the Boltzmann transport equation (BTE) with a full phonon dispersion model (FPDM) to capture mode-resolved velocities, scattering processes, and nonequilibrium phonon populations in silicon. The analysis demonstrates that the increase in TR near NHs is not caused by the long MFP itself but by the low specific heat of long-MFP phonons that do not scatter directly with optical modes. These phonons heat readily when energy is supplied, steepening the local temperature gradient near the NH and thereby enhancing TR. By resolving the spectral contributions to the phonon transport resistance and temperature gradients, we identify the critical role of the modal specific heat in nonlocal phonon transport. These results provide new physical insights into nanoscale thermal management and highlight the importance of spectral mode resolution in modeling heat dissipation in electronic devices.

---

[a] E-mail : jinjs@cst.ac.kr

# 1 Introduction

A considerable amount of effort has been devoted to understanding phonon transport near nanoscale hotspots, as this knowledge offers pathways for phonon engineering strategies in the thermal management of advanced electronic, optoelectronic, and quantum nanodevices [1-3]. Previous theoretical and experimental studies have consistently reported that the thermal resistance (TR) around hotspots is significantly greater than that predicted by Fourier-based diffusive transport models [1,4]. This inefficiency in heat dissipation has been attributed to two main factors [2]: (1) phonon transport near nanoscale hotspots (NHs) involves ballistic transport (BT), and (2) electron–phonon scattering processes induce selective phonon excitation (SE), where thermally excited electrons predominantly transfer their energy to optical phonon modes (OMs) with very low group velocities, and the energy stored in this OM is subsequently and selectively funneled into high-frequency acoustic phonon modes (AMs). These modes act as bottlenecks in the overall thermal transport process [5,6].

Importantly, V. Chiloyan et al. [7] suggested that in silicon at 300 K, under nonthermal conditions without phonon scattering, long mean free path (MFP) phonons emerging from the hotspot can yield higher heat transfer rates than those predicted by Fourier analysis. This raises the need to revisit the common assumption that reduced phonon scattering near NHs necessarily leads to enhanced TR. In practice, phonons emitted from a hotspot inevitably undergo scattering, with background phonons of various frequencies already present at a given temperature.

BT is traditionally associated with situations where the transport length is comparable to the phonon MFP. However, it can also occur when the characteristic length scale of the temperature gradient is of similar magnitude [8,9]. Because phonon MFPs span a broad spectral range [9,10], the same temperature gradient may appear long to short-MFP modes but short to long-MFP modes, resulting in distinct sensitivities to nonlocal effects. This spectral dependence highlights the importance of mode-resolved analysis and motivates the introduction of a new spectral parameter, $-\nabla T/\Lambda_f$ (here, $\nabla T$ is the temperature gradient, and $\Lambda_f$ denotes the phonon MFP in a thin film), to capture this issue. Considering these mechanisms, regions near the NH inherently host a variety of background phonon modes set by the local temperature, and emitted phonons, including long-MFP modes, dissipate energy through scattering among the different frequency phonon modes. Additionally, to fully account for SE, it is essential to model phonon transport near the NH in semiconductor devices by incorporating interactions between phonons of different energies or frequencies.

A precise understanding of phonon transport near the NH requires explicit consideration of both phonon dispersion and polarization. While previous studies have focused primarily on polarization [2, 3], this work employs the full phonon dispersion model (FPDM) within the Boltzmann transport equation (BTE) framework to incorporate both effects in phonon scattering across energy states. Using silicon as a model system, we compute the TR induced by a 10 nm NH in the Si layer of a silicon-on-insulator (SOI) transistor, analyze its impact on spectral phonon modes and the spectral temperature gradient, and identify the origin of the additional resistance generated near the hotspot.

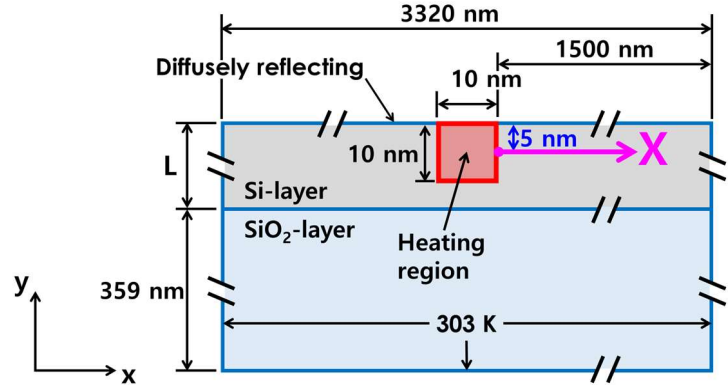

Fig. 1 Computational domain used in this study, identical to that presented in Ref. [13], which was constructed based on the experimental geometry of Goodson et al. [14]

## 2 Methodology

First-principles (FP) simulations provide reliable spectral properties, such as phonon relaxation times, but their direct integration with device models remains infeasible owing to the prohibitive computational cost. In practice, FP-derived data are often used as inputs for semiconductor simulations, yet this approach lacks a rigorous definition of scattering rates between phonons of different frequencies or energies. As a result, only effective spectral data, capturing dispersion and polarization but not explicit inter- or intra-branch scattering, are available.

In this study, phonon transport near the NH is modeled via the BTE-based FPDM under the relaxation-time approximation, with mode-resolved velocities and relaxation times obtained from experimental dispersion relations and perturbation theory [5, 11]. The FPDM includes longitudinal, transverse, and optical branches, resolving intra- and interbranch scattering under energy and momentum conservation. The impurity, boundary, and three-phonon normal and Umklapp processes are included. This framework recovers Fourier diffusion in the macroscopic limit while resolving nonequilibrium phonon populations critical to nanoscale heat transport in silicon [11].

The volumetric energy density per unit solid angle for a given frequency band $i$ is defined as $e_i'' = \int_{\Delta \omega i} \hbar \omega f_\omega D(\omega) d\omega$, where $\hbar$ is the reduced Planck's constant, $\omega$ is the phonon angular frequency, $\Delta \omega$ is the frequency width, $f_\omega$ is the phonon distribution function, and $D(\omega)$ is the phonon density of states. A steady-state analysis of BTE-based FPDM for the AM is given as [5]

$$\nabla \cdot \left(v_{g,i}\hat{s}e_i''\right) = \left(e_i^0 - e_i''\right)\gamma_{ii} + \sum_{\substack{j=1 \\ j \neq i}}^{N_{bands}} \left\{\left(\frac{1}{4\pi}\int_{T_{ref}}^{T_{ij}} C_i dT - e_i''\right)\gamma_{ij}\right\}, \tag{1}$$

where $v_g$ is the spectral phonon group velocity, $T_{ref}$ is the reference temperature ($T_{ref}$ = 303 K in this study, Fig. 1), $\hat{s}$ is the phonon propagation direction defined by the polar and azimuthal angles [5, 11], and the equilibrium energy density, $e_i^o$, is given for the $i$-th frequency band:

$$e_i^o = \frac{1}{4\pi}\int_{4\pi} e_i'' d\Omega = \frac{1}{4\pi}\int_{T_{ref}}^{T_i} C_i dT = \frac{e_i}{4\pi}, \tag{2}$$

where the index $j$ denotes either acoustic or optical phonon scattering with the $i$-th band. The BTE-based FPDM for the OM is also given as [5]

$$\frac{\partial e_o}{\partial t} = \sum_{j=1}^{N_{bands}-1}\left(\int_{T_{ref}}^{T_{oj}} C_o dT - e_o\right)\gamma_{oj} + q_{vol} \tag{3}$$

where $C_i$ and $C_o$ are the specific heats of the $i$-th acoustic and optical phonon bands, respectively, and $e_o$ is the volumetric energy density for an optical band. $T_{ij}$ and $T_{oj}$ are the interaction temperatures given as (an example for $T_{ij}$):

$$\int_{T_{ref}}^{T_{ij}}\left(\frac{C_i}{\Delta\omega_i} + \frac{C_j}{\Delta\omega_j}\right)dT = \int_{T_{ref}}^{T_i}\frac{C_i}{\Delta\omega_i}dT + \int_{T_{ref}}^{T_j}\frac{C_j}{\Delta\omega_j}dT \tag{4}$$

In Eqs. (1) and (3), $\gamma_{ij}$, $\gamma_{ii}$ and $\gamma_{oj}$ are the scattering rates for each scattering process, and $N_{bands}$ is the total number of frequency bands ($N_{bands} = N_{LA} + N_{TA} + N_{optical}$ with $N_{LA} = N_{TA} = 6$ and $N_{optical} = 1$, as given in Refs. [5, 11]). Here, LA and TA denote the longitudinal and transverse acoustic phonon modes, respectively. The detailed calculations of these quantities, including the phonon specific heats ($C_i$ and $C_o$), are given in Refs. [5, 11]. In the present simulation, the phonon specific heats are taken as temperature independent since they remain nearly constant above 300 K [11]. The source term $q_{vol}$ in Eq. (3) denotes the volumetric heat generation, i.e., the total energy transferred from electrons to phonons through electron–phonon scattering processes. In this study, $q_{vol} = 10^{18}$ W/m³ is chosen. To calculate the lattice temperature, $T_L$, the total phonon energy ($e_{total}$) is defined as

$$e_{total} = e_o + \sum_{i=1}^{N_{bands}-1} e_i = \int_{T_{ref}}^{T_L} C_{total} dT,$$ where $C_{total}$ is the total specific heat of the solid. The quantitative values of the phonon–phonon scattering rates ($\gamma_{ij}$, $\gamma_{ii}$ and $\gamma_{oj}$) between different bands are reported in Ref. [5], and these values were adopted in the present calculations.

Accurate evaluation of the phonon transport TR via FPDM requires modal heat flux calculations. Since FPDM neglects the directional dependence of phonon group velocities, errors arise during solid-angle integration. Although this approximation may be acceptable for total heat flux under gentle temperature gradients, it becomes unsuitable in regions with sharp thermal gradients, such as near NHs, where local modal heat flux must be accurately resolved.

We analyze the role of spectral phonon modes in determining the local TR near nanoscale hotspots. In particular, the effect of spectral temperature gradients on the phonon spectral TR in the vicinity of NHs is evaluated. Assuming one-dimensional transport along the Si layer (a validation for this, will be discussed later, Fig. 4.) we adopt Majumdar's spectral heat flux model [12] to compute the spectral TR, $R_\omega = \Delta T/q_{x,\omega}$ with the temperature drop $\Delta T$ and the spectral heat flux

$$q_{x,\omega} = \int_{\Delta \omega i} v_g \hbar \omega f_{\omega,\text{Eq}}(x) D(\omega) d\omega, \tag{5}$$

where $f_{\omega,\text{Eq}}(x) = f_\omega^0(x) - \tau v_g \dfrac{df_\omega^0}{dT}\dfrac{dT}{dx}$ with $\dfrac{df_\omega^0}{dT} = \hbar\omega \exp\left(\dfrac{\hbar\omega}{k_B T}\right) / k_B T^2 \left[\exp\left(\dfrac{\hbar\omega}{k_B T}\right) - 1\right]^2$ via the equilibrium Bose–Einstein distribution function of $f_\omega^0 = \left[\exp\left(\dfrac{\hbar\omega}{k_B T}\right) - 1\right]^{-1}$, where $k_B$ is Boltzmann's constant. This approach evaluates the ratio of the spectral heat flux to the modal temperature difference, with the assumption that each phonon mode can be assigned an equilibrium temperature. Phonon modes near NHs may deviate from equilibrium, yet in this study, each mode is characterized by a modal temperature obtained from the BTE solution. This effective (or *modal*) equilibrium assumption simplifies the analysis while capturing the essential role of modal temperatures in phonon transport, but it should be recognized as a limitation of the present approach.

For the LA and TA bands, a higher band number corresponds to a higher frequency; Number 6 corresponds to the band with the highest frequency, whereas number 1 corresponds to the lowest frequency. With respect to the spectral specific heat ($C_\omega$), $C_\omega$ is proportional to the phonon frequency ($\omega$) because $C_\omega \sim u \sim \hbar\omega$, where $u$ is the internal energy of a solid. Therefore, it is important to note that higher specific heats are associated with higher band numbers in each branch. $q_{vol}$ is partitioned between AM and OM; the amount transferred to AM is denoted by $q_a$. When a nonzero portion of $q_{vol}$

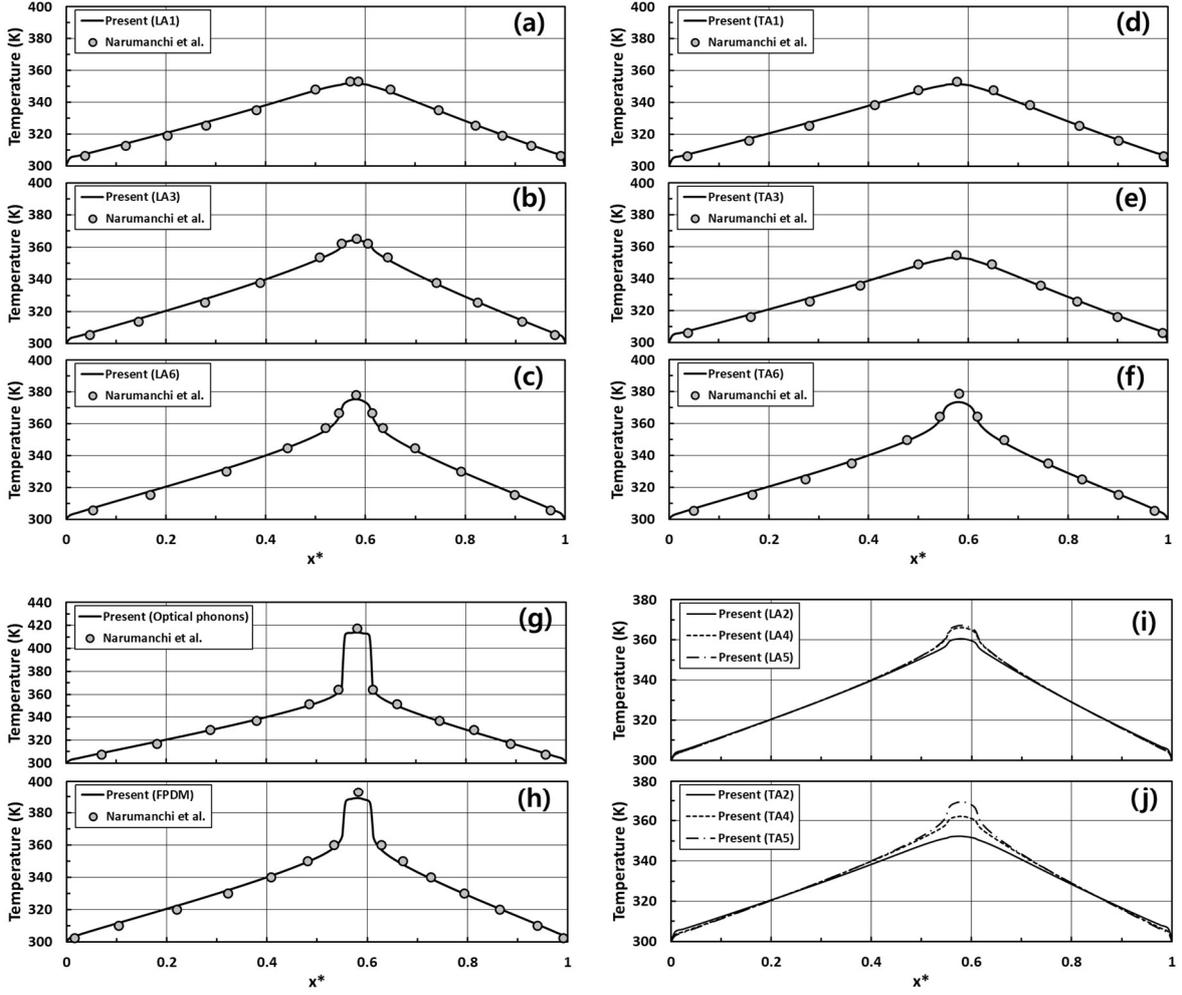

Fig. 2 Comparison of calculated temperature distributions with Narumanchi et al. [5], showing good agreement within error margins: (a) LA1, (b) LA3, (c) LA6, (d) TA1, (e) TA3, (f) TA6. Figures (g)–(j) show only our results, since Narumanchi et al. did not report results for these cases

is allocated to AM ($q_a \neq 0$), the heat source is uniformly distributed among the twelve acoustic bands, with each receiving $q_{vol}/12$.

The computational domain used in this study (Fig. 1) is identical to that presented in Ref. [13], which was constructed on the basis of the experimental geometry of Goodson et al. [14]. Localized hot spots generated by strong electron–phonon or photon–phonon interactions produce high-energy nonequilibrium phonons. These phonons relax through anharmonic decay, forming a cascade in which the OM sequentially decays into lower-energy AM. The efficiency of this relaxation chain depends on the hotspot size: in small hotspots, spatial confinement restricts the phase space and enforces a cascade-like energy flow [15], whereas in large hotspots, the expanded phase space activates multiple scattering channels, enabling rapid thermalization through direct transitions into low-energy modes [16]. To capture nonequilibrium phonon dynamics under strong confinement, the hotspot size in this study is fixed at $10 \times 10$ nm$^2$. In the simulation of the computational domain shown in Fig. 1, the BTE-based FPDM is employed in the Si layer, and the Fourier diffusion equation is solved in the SiO$_2$ region [5]. The spatial grids are chosen as 130 (in the $x$ direction) × 64 (in the $y$ direction) for a Si layer thickness (L) of L = 41 nm, 130 × 70 for L = 78 nm, and 130 × 76 for L = 177 nm. An angular resolution in the octant of 6 × 6 is used. These grids provide converged results within 0.1% regardless of the mesh size.

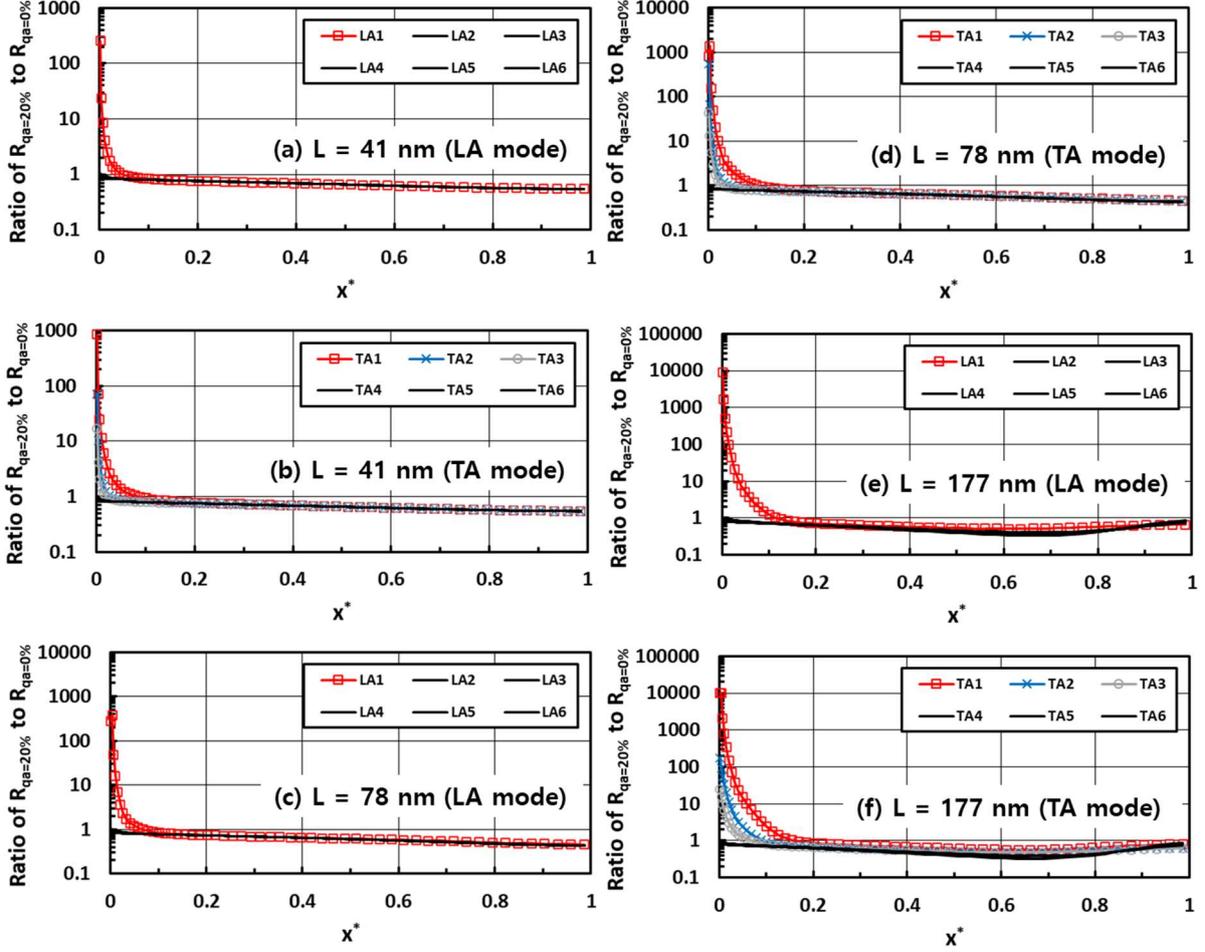

Fig. 3 Ratio of thermal resistance for $q_a = 20\%$ relative to $q_a = 0\%$ across LA and TA branches. LF modes exhibit a sharp increase near the hotspot due to their long mean free paths.: (a) L = 41 nm (LA mode), (b) L = 41 nm (TA mode), (c) L = 78 nm (LA mode), (d) L = 78 nm (TA mode), (e) L = 177 nm (LA mode), (f) L = 177 nm (TA mode)

## 3 Results and discussion

According to previous studies, 20.4% of the thermal energy ($q_{vol}$) generated by excited electrons is transferred to AM, while the remaining 79.6% is transferred to OM. [2, 6]. Furthermore, the energy delivered to AM is distributed nearly uniformly across different phonon modes in the Si layer [3]. In this study, we therefore assume that 20% of the thermal energy generated by excited electrons is transferred to AM, whereas the remaining 80% is delivered to OM, with the AM energy equally distributed among six phonon bands. For the estimation of TR via Eq. (5), the accuracy of the simulated temperature distributions is critical. Their reliability was assessed via comparison with the numerical results of Narumanchi et al. [5] for the case of $q_a = 0$, as shown in Fig. 2. The comparison shows good agreement within acceptable error margins. Although the SOI transistor geometry in Ref. [5] differs from that in Fig. 1, the consistent temperature profiles affirm the robustness of our results.

Figure 3 shows the ratio of TR for $q_a = 20\%$ relative to $q_a = 0\%$ for the longitudinal acoustic (LA1~LA6) and transverse acoustic (TA1~TA6) branches. The position $x^* = 0$ corresponds to the edge of the hotspot region, as marked X in Fig. 1, and $x^* = X/1500$ nm. For the low-frequency phonon (LF) modes (LA1 and TA1, TA2, TA3), the TR increases sharply near the NH. These modes have relatively long MFPs, in line with previous observations that phonons with disproportionately long MFPs relative to the hotspot size fail to efficiently transport heat away from the NH [1, 4].

Phonon transport in the Si layer is known to occur predominantly along the *x*-axis due to the very low thermal conductivity of the underlying SiO$_2$ layer [5]. As expected, a notable exception arises immediately beneath the hotspot, where the thin Si layer allows substantial *y*-directional transport. To validate the assumption of one-dimensional heat transport, $T_L$ distributions were compared between the bottom and top of the Si layer for $q_a = 0\%$ and $q_a = 100\%$, as shown in Fig. 4. For all the cases of $L = 41$ nm, 78 nm, and 177 nm, the $T_L$ is lower for $q_a = 100\%$ than for $q_a = 0\%$, as expected. With increasing layer thickness $L$, the overall $T_L$ decreases further, and the top-surface temperature profile, where the hotspot is located, exhibits a markedly steeper decline, whereas the bottom-surface profile shows a more gradual variation due to its distance from the hotspot. The maximum difference between the top and bottom surfaces was observed for $L = 177$ nm at $q_a = 0\%$, but even in this case, it was less than 1.6%, indicating that although *y*-directional transport is locally enhanced just beneath the hotspot, the one-dimensional phonon transport assumption remains justified throughout the Si layer. Another observation is that although the comparison between $q_a = 0\%$ and $q_a = 100\%$ may seem intuitive, supplying $q_{\text{vol}}$ to the AM reduces the $T_L$ for all Si layer thicknesses due to the high group velocity of AM.

To elucidate the physical mechanisms underlying NH phonon transport, Fig. 5 [(a), (c), (d)] shows the variation in $-\nabla T/\Lambda_f$ with $x^*$ for LF modes, comparing $q_a = 0\%$ and $q_a = 20\%$. Here, $\Lambda_f$ was estimated using $\Lambda_f = \phi \Lambda_b$, where $\Lambda_b$ is the bulk phonon MFP, and the suppression function $\phi$ was determined on the basis of the approach outlined in Ref. [17] for in-plane thermal transport in silicon thin films. Boundary scattering was assumed to be fully diffuse, corresponding to a specularity parameter of zero.

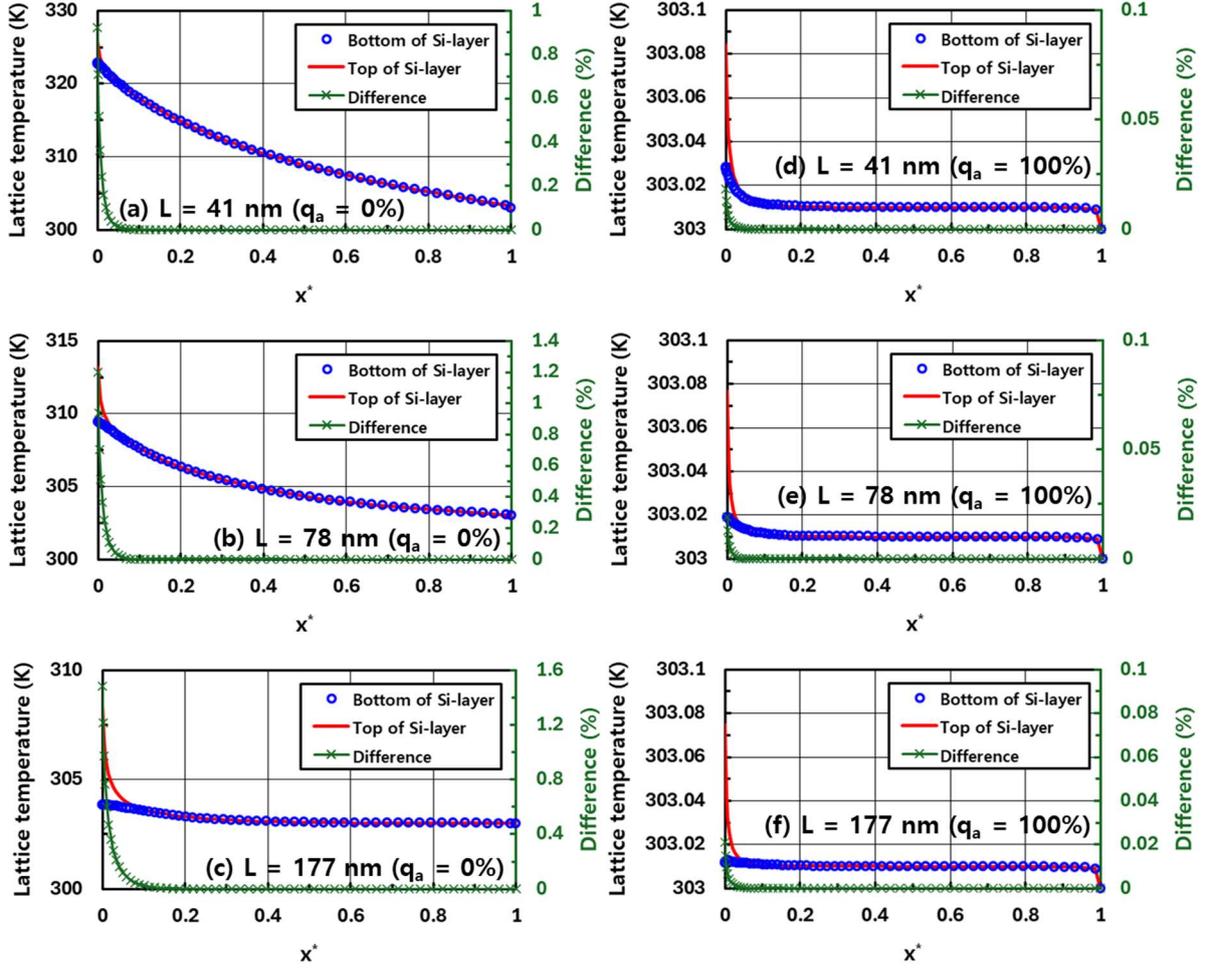

Fig. 4 Lattice temperature distributions as a function of $x^*$ in the Si-layer for $q_a$ = 0% and 100%, validating the one-dimensional phonon transport assumption: (a) $L$ = 41 nm ($q_a$ = 0%), (b) $L$ = 78 nm ($q_a$ = 0%), (c) $L$ = 177 nm ($q_a$ = 0%), (d) $L$ = 41 nm ($q_a$ = 100%), (e) $L$ = 78 nm ($q_a$ = 100%), (f) $L$ = 177 nm ($q_a$ = 100%)

For $L$ = 41 nm at $q_a$ = 0%, Narumanchi et al. [5] reported that LF modes, with frequencies much lower than those of OM (e.g., the highest-frequency TA3 among LF modes is 2.2 THz versus 14.3 THz for OM), cannot satisfy momentum and energy conservation simultaneously, resulting in negligible LF–OM scattering with a scattering rate ≈ 0. As a result, LF modes contribute little to energy transport near the NH, resulting in a low $-\nabla T/\Lambda_f$ at $x^*$ = 0. As they propagate, however, they gradually gain energy through scattering with AM that strongly couples to OM, which causes $-\nabla T/\Lambda_f$ to increase before decreasing again owing to energy dissipation at the boundary ($x^*$ = 1). The TR follows the same trend, as it is proportional to $-\nabla T/\Lambda_f$.

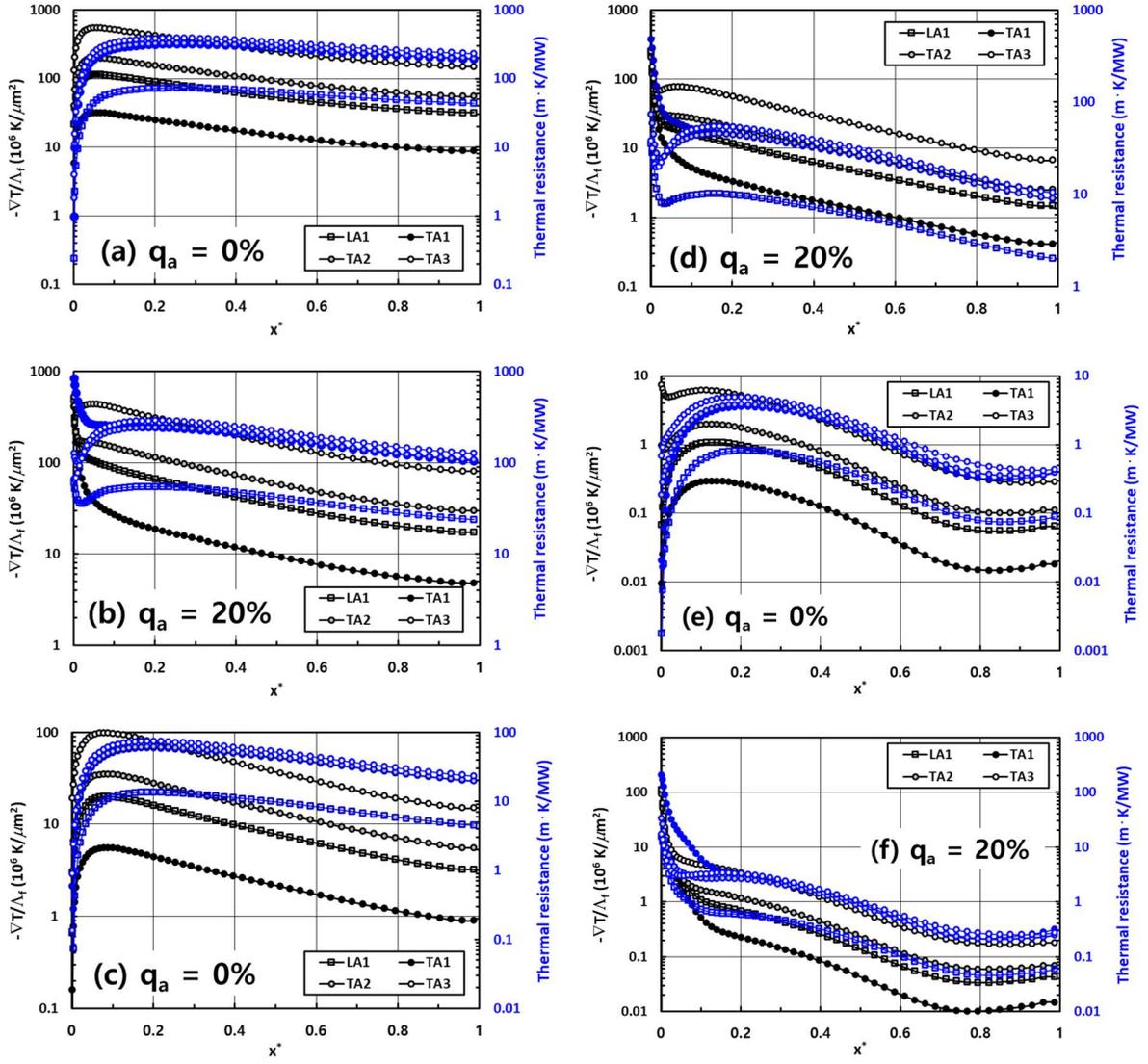

Fig. 5 Distributions of $-\nabla T/\Lambda_\mathrm{f}$ for LF modes at $q_a = 0\%$ [(a), (c), and (e)] and 20% [(b), (d), and (f)] for $L$ = 41, 78, and 177 nm. LF modes show enhanced thermal resistance at $x^* = 0$ under $q_a = 20\%$. A distinct trend is observed for TA3 due to its coupling with selected LA modes at $L$ = 177 nm (the underlying mechanism is explained in the main text): (a) and (b) for $L$ = 41 nm, (c) and (d) for $L$ = 78 nm, (e) and (f) for $L$ = 177 nm

As shown in Fig. 5, $-\nabla T/\Lambda_\mathrm{f}$ is plotted as a function of $x^*$ for $L$ = 41, 78, and 177 nm at $q_a = 0\%$ and 20%. For each $L$, a clear difference appears. With heat supplied from the NH to the AM ($q_a = 20\%$), $-\nabla T/\Lambda_\mathrm{f}$ and the TR increase near the NH ($x^* = 0$) with increasing $L$ and then decrease as $x^*$ increases. The key result is that, for LF modes, the TR at $x^* = 0$ increases substantially under $q_a = 20\%$, relative to $q_a = 0\%$. In contrast, for high-frequency (HF) phonon modes (Fig. 6), there is almost no change. This arises because phonon modes with low specific heats (e.g., LF modes) heat readily when energy is supplied but cool rapidly via phonon scattering, producing a sharp temperature drop. Thus, this process leads to the enhanced TR observed for LF modes.

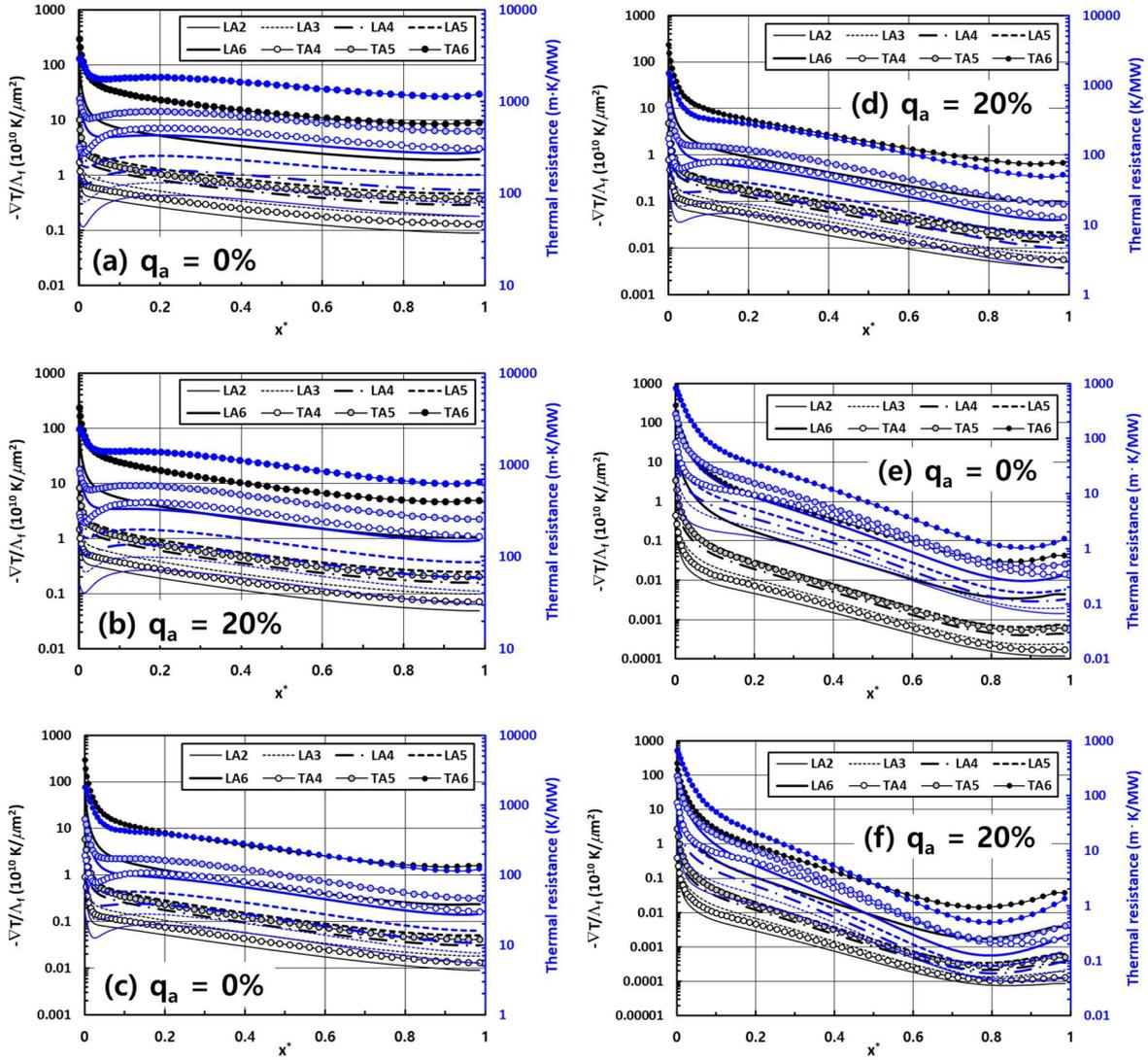

Fig. 6 – $\nabla T/\Lambda_f$ distributions for HF modes at $q_a$ = 0% [(a), (c), and (e)] and 20% [(b), (d), and (f)] for $L$ = 41, 78, and 177 nm. Unlike LF modes, HF modes show nearly identical results for both $q_a$ values, with slightly lower resistance at $x^* = 1$ under $q_a = 20\%$: (a) and (b) for $L$ = 41 nm, (c) and (d) for $L$ = 78 nm, (e) and (f) for $L$ = 177 nm

However, TA3 in Fig. 5(e) shows a distinct trend. The TA modes emerging from the NH cannot satisfy energy and momentum conservation simultaneously, so direct TA–TA scattering is negligible. In contrast, several LA modes scatter strongly with OM, notably LA3 (66.7%), LA4 (52.2%), and LA6 (57.1%), where the percentages indicate the fraction of their total scattering rate that involves OM.

Unlike other LF modes, TA3 couples more strongly to these LA modes and receives thermal energy from the NH via them. This transfer is suppressed by strong surface confinement but becomes evident once confinement is relaxed. The relatively long mean free paths of LA3 (~187 nm) and LA4 (~131 nm) at 300 K allow efficient energy transfer to TA3 when $L$ = 177 nm, whereas LA6 (~46 nm) still contributes partly to its higher frequency. As a result, $-\nabla T/\Lambda_f$ for TA3 starts higher than those for the other LF modes, decreases, and then slightly increases before decreasing again.

For $q_a$ = 20% [Fig. 5(b), (d), (f)], LF modes directly acquire one-fifth of the injected energy, producing a large $-\nabla T/\Lambda_f$ at $x^*$ = 0 that decays with distance and yields a maximum TR near the hotspot. With increasing Si layer thickness, the surface confinement of long-MFP phonons weakens, increasing their transport and increasing these trends. In this regime, both $-\nabla T/\Lambda_f$ and TR decrease with $x^*$, and their decay becomes increasingly similar as the thickness increases. Once the Si layer is sufficiently thick ($L$ = 177 nm) or when more energy is transferred to acoustic phonons, the phonon-mediated internal resistance diminishes, and boundary effects dominate, causing TR to rise again near $x^*$ = 1.

For completeness, Fig. 6 compares the $-\nabla T/\Lambda_f$ distributions of the HF modes at $q_a$ = 0% and 20% for $L$ = 41, 78, and 177 nm. In contrast to the LF modes in Fig. 5, direct comparison of $q_a$ = 0% and 20% shows that the results are nearly identical across all Si layer thicknesses ($L$ = 41, 78, and 177 nm), with only a slight reduction in TR at $x^*$ = 1 for $q_a$ = 20% owing to the higher group velocity of AM, which enables more efficient energy transfer.

A comparison of Figs. 5 and 6 for each Si layer thickness reveals that when heat is supplied to the AM, the TR of the HF modes is greater than that of the LF modes. As a result, heat from the NH is dissipated mainly through LF modes. Since the TR of LF modes increases substantially under $q_a$ = 20% relative to $q_a$ = 0%, heat dissipation from the NH becomes much less efficient than that predicted by Fourier analysis (the diffuse assumption).

To examine the behavior of LF modes in more detail, Figs. 7–9 present $-\nabla T/\Lambda_f$ distributions for various $q_a$ values in the near-hotspot region ($0 \leq x^* \leq 0.1$). As explained earlier, $-\nabla T/\Lambda_f$ and the TR at $x^*$ = 0 for LF modes increase with $q_a$, as shown for each $L$. The decay with distance becomes more rapid at larger $q_a$. These results clearly demonstrate that the low specific heat of LF modes plays a key role in enhancing TR near the NH.

## 4 Conclusions

For decades, the enhanced thermal resistance observed near nanoscale hotspots has been attributed to long mean free path phonons failing to couple efficiently to the hotspot because their mean free paths

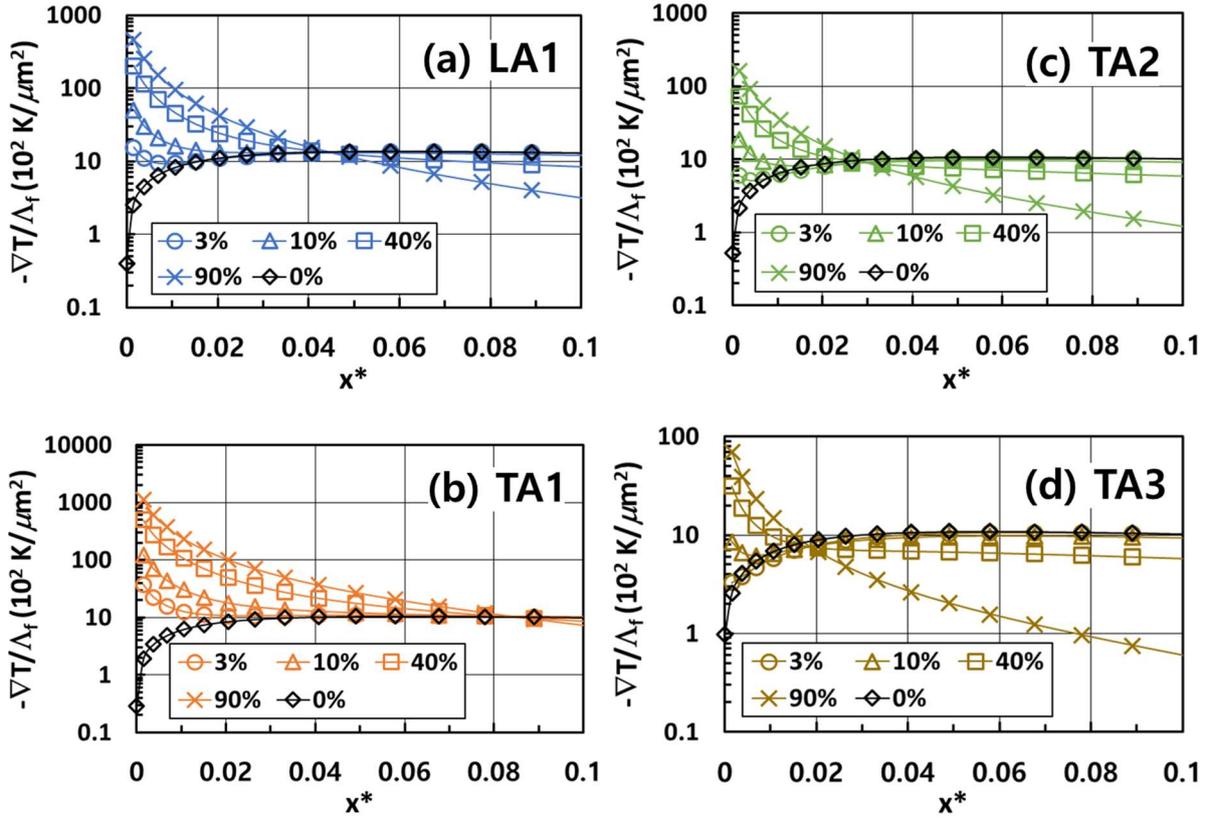

Fig. 7 $-\nabla T/\Lambda_f$ distributions of LF modes near the hotspot ($0 \leq x^* \leq 0.1$) for various $q_a$ values at a Si-layer thickness of $L$ = 41 nm: (a) LA1, (b) TA1, (c) TA2, (d) TA3

far exceed its size. The present analysis shows, however, that the true origin lies in the low specific heat of long-mean free path phonons that do not scatter directly with optical modes. Owing to their low specific heat, these phonons experience a rapid temperature rise when energized, which steepens the temperature gradient by scattering with other acoustic phonon modes near the hotspot and thereby increases the thermal resistance at the nanoscale hotspots.

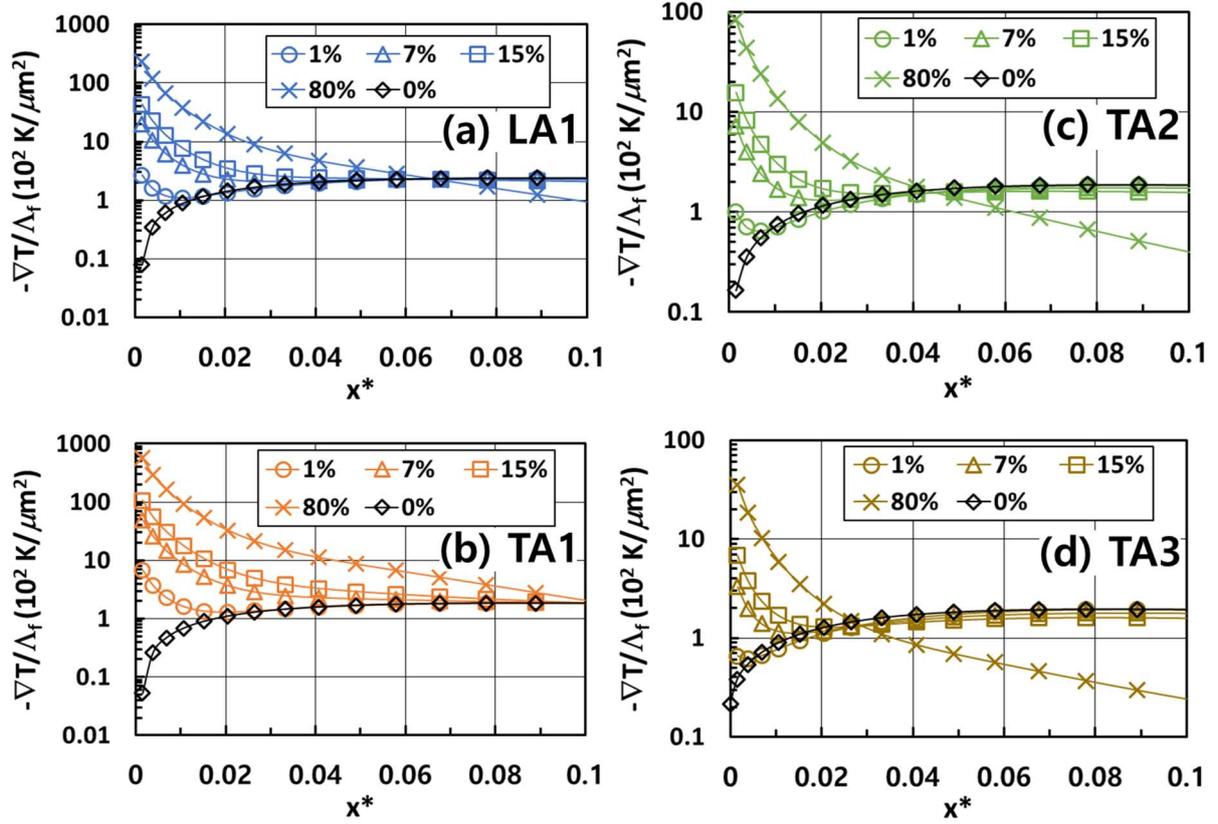

Fig. 8 $-\nabla T/\Lambda_f$ distributions of LF modes near the hotspot ($0 \leq x^* \leq 0.1$) for various $q_a$ values at a Si-layer thickness of $L$ = 78 nm: (a) LA1, (b) TA1, (c) TA2, (d) TA3

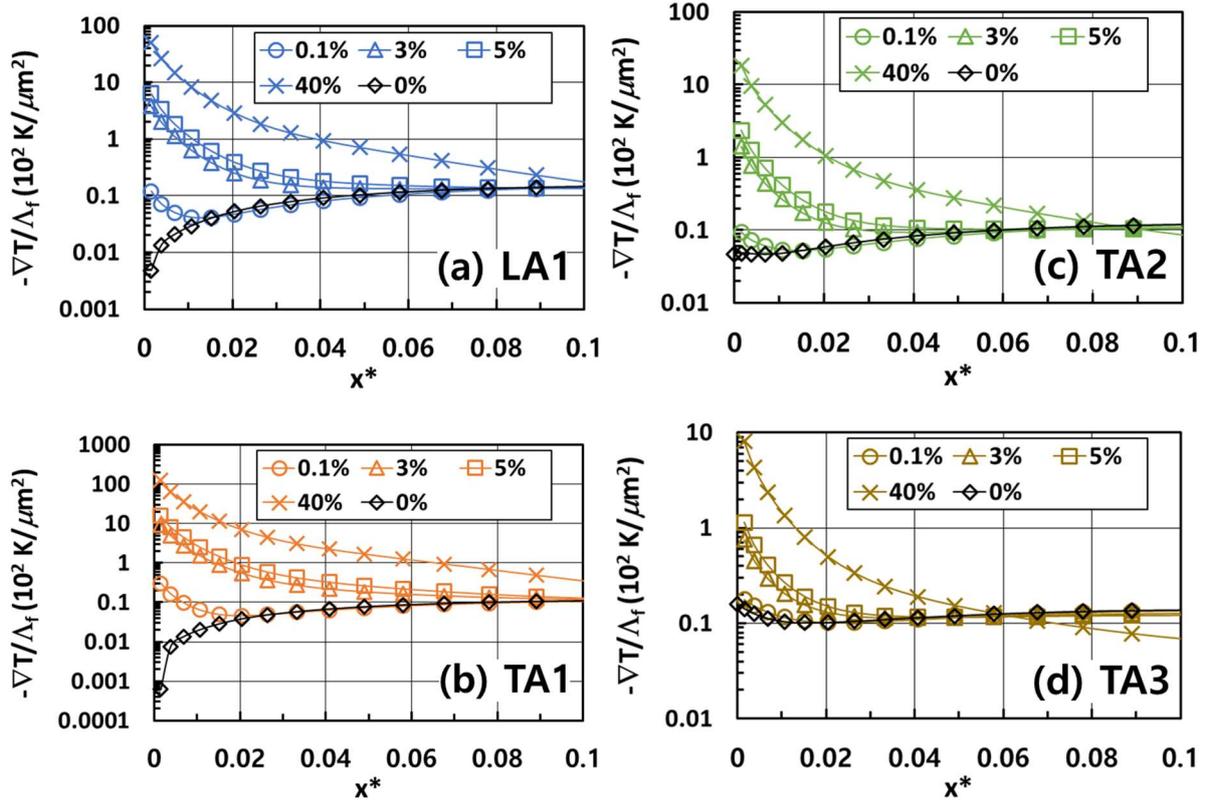

Fig. 9 $-\nabla T/\Lambda_\text{f}$ distributions of LF modes near the hotspot ($0 \leq x^* \leq 0.1$) for various $q_a$ values at a Si-layer thickness of $L$ = 177 nm: (a) LA1, (b) TA1, (c) TA2, (d) TA3